\documentclass{ws-ijmpcs}
\usepackage{txfonts}
\usepackage{amssymb}

\begin{document}

\markboth{Haoqi Lu} {Reactor antineutrino experiment}

%
\catchline{}{}{}{}{}
%

\title{Reactor antineutrino experiments}

\author{Haoqi Lu}

\address{Institute of High Energy Physics, Chinese Academy of Sciences \\
Yuquan Road 19-B, Beijing 100049,P.R. China\\
Member of Daya Bay collaboration\\
 luhq@ihep.ac.cn}

\maketitle

\begin{history}
\received{Day Month Year}
\revised{Day Month Year}
\end{history}

\begin{abstract}
Neutrinos are elementary particles in the standard model of particle
physics. There are 3 flavors of neutrinos that oscillate among
themselves. Their oscillation can be described by a 3$\times$3
unitary matrix, containing three mixing angles $\theta_{12}$,
$\theta_{23}$, $\theta_{13}$, and one CP phase. Both $\theta_{12}$
and $\theta_{23}$ are known from previous experiments. $\theta_{13}$
was unknown just two years ago. The Daya Bay experiment gave the
first definitive non-zero value in 2012. An improved measurement of
the oscillation amplitude $\sin^{2}2(\theta_{13})$ =
$0.090^{+0.008}_{-0.009}$ and the first direct measurement of the
$\bar\nu_{e}$ mass-squared difference $\mid$$\Delta m^2_{ee}$$\mid$
= $(2.59^{+0.19}_{-0.20})\times10^{-3} \rm eV^{2}$ were obtained
recently. The large value of $\theta_{13}$ boosts the next
generation of reactor antineutrino experiments designed to determine
the neutrino mass hierarchy, such as JUNO and RENO-50 .

\end{abstract}

\ccode{PACS numbers:} 14.60.Pq, 29.40.Mc, 13.15.+g

\section{Introduction}

The neutrino is a fundamental particle and was first discovered in
1956 by Cowan and Reines \cite{firstNu}$^{,}$\cite{firstNu2}. In the
last few decades, it has been proven that the observed neutrino
oscillations can be described in a 3-flavor neutrino framework. A
parameterization of the standard Pontecorvo-Maki-Nakagawa-Sakata
(PMNS) matrix describing the unitary transformation relating the
mass and flavor eigenstates, defines the three mixing angles
($\theta_{23}$, $\theta_{12}$, and $\theta_{13}$) and one
charge-parity(CP)-violating phase \cite {PMNS}$^{,}$\cite {PMNS2}.
$\theta_{12}$ is about 34$^{\textordmasculine}$ and determined by
solar and reactor neutrino experiments. $\theta_{23}$ is about
45$^{\textordmasculine}$ and determined by atmospheric and
accelerator neutrino experiments. An upper limit of the last unknown
angle $\theta_{13}$ was given by CHOOZ
$\sin^{2}(2\theta_{13})$$<$0.15 at 90\% confidence level
 (C.L.)\cite{ChooZ}. It was hinted to be non-zero by recent results
 from the T2K \cite{T2K}, MINOS \cite{MINOS} and Double Chooz
experiments\cite{Double Chooz}. The value of $\theta_{13}$ will
guide the designs of future experiments for the measurements of the
mass hierarchy and CP-violation.

 Long-baseline accelerator experiments have limited sensitivity to the $\theta_{13}$ mixing angle due to
dependencies on the yet unknown mass hierarchy and CP-violating
phase. Reactor antineutrino experiments can provide a clear and
accurate measurement of $\theta_{13}$, due to their pure
antineutrino source, clear
    signal and independence of the CP phase and matter effects. For reactor-based antineutrino experiments, $\theta_{13}$ can be
determined in terms of the survival probability of $\bar{\nu_{e}}$
at certain distances from the reactors,
\begin{equation}
P(\bar \nu_{e}\rightarrow \bar \nu_{e}) = 1 - \cos^4\theta_{13}
\sin^{2}2\theta_{12} \sin^2 \Delta_{21}
-\sin^22\theta_{13}(\cos^2\theta_{12} \sin^2\Delta_{31} + \sin^2
\theta_{12} \sin^2\Delta_{32}) \label{diseq}
\end{equation}

where $\Delta_{ji}$ $\equiv$ 1.267$\Delta^{2}_{ji}(\rm eV^{2})$
L(m)/E(MeV), E is the $\bar\nu_{e}$ energy in MeV and L is the
distance in meters between the $\bar \nu_{e}$ source and the
detector. $\Delta m^{2}_{ji}$ is the difference between the
mass-squares of the mass eigenstates $\nu_{j}$ and $\nu_{i}$(
$\Delta m^{2}_{ji}$ =  $\Delta m^{2}_{j}$ -  $\Delta m^{2}_{i}$).
Since $\Delta m^{2}_{21}$ $\ll$ $\mid$ $\Delta
m^{2}_{31}$$\mid$$\approx$$\mid$$\Delta$
$m^{2}_{32}$$\mid$\cite{PDG}, the short distance ($\sim$ km) reactor
$\bar \nu_{e}$ oscillation is mainly determined by the $\Delta_{3i}$
terms.

\begin{figure}[pb]
\centerline{\includegraphics[width=6.0cm]{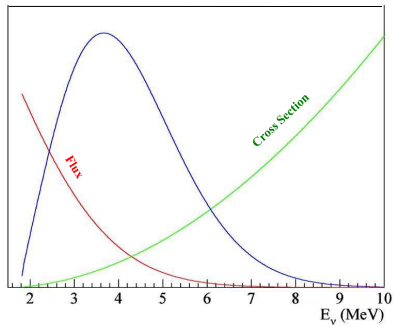}}
\vspace*{8pt} \caption{The observable $\bar\nu$ spectrum, a product
of the antineutrino flux from reactors and cross section of IBD
reaction. \label{nu_spectrum}}
\end{figure}

   Most reactor antineutrino oscillation experiments measure antineutrino
events via the inverse beta decay(IBD) reaction $\bar{\nu} + p \to
e^{+} + n $. The IBD reaction is characterized by two time
correlated events, the prompt signal coming from the production and
subsequent annihilation of the positron, and the delayed signal from
the capture of the neutron in the liquid scintillator. For the Daya
Bay experiment, 0.1\% gadolinium (Gd)-doped liquid scintillator is
used to increase the capture cross section of thermal neutrons on Gd
and reduce the capture time (about 30 $\rm\mu$s) to suppress
accidental coincidence backgrounds. The total energy of gammas from
neutron capture on Gd is about 8 MeV, which is much higher than the
energy range of background radioactivity. The observable
antineutrino spectrum is shown in Fig.\ref{nu_spectrum}.

Due to the small value of $\theta_{13}$, precise measurements are
required to reduce the experimental uncertainties. In previous
reactor antineutrino experiments, such as Palo Verde \cite{Palo
Verde} and CHOOZ \cite{ChooZ}, single detectors were utilized for
antineutrino detection. Their measurements had large uncertainty
from errors related to detector response and antineutrino flux
predictions. There are three ongoing reactor experiments : Daya Bay,
Double Chooz, and RENO. They use multiple detectors at different
baselines to reduce correlated uncertainties by a relative
measurement. A comparison of the three experiments is shown in Table
1. The baselines of their far detectors are 1.65, 1.05 and 1.44km,
respectively. \cite{JunTalk}.

The following content of this paper mainly focuses on the Daya Bay
experiment and future prospects of reactor antineutrino experiments.

\begin{table}
\tbl{Comparison of three reactor experiments }
{\begin{tabular}{@{}ccccc@{}} \toprule Exp. & Power &
Det.(t) & Overburden & 3y Sens. \\
& (GW) & (N/F)& (m.w.e)N/F &(90\%CL)\\ \colrule
Double Chooz& 8.5 &8/8& 120/300& 0.03\\
RENO& 16.5 &16/16& 120/450 &0.02 \\
DYB& 17.4& 80/80& 250/860& 0.008\\
 \botrule
\end{tabular} \label{ta1}}
\end{table}

\begin{figure}[pb]
\centerline{\includegraphics[width=7.0cm]{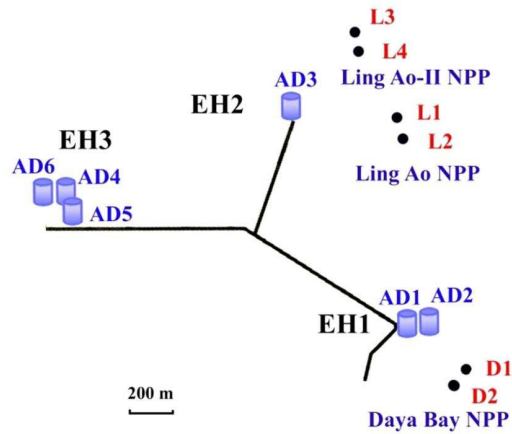}}
\vspace*{8pt} \caption{Layout of the Daya Bay experiment. There are
six reactor cores (D1, D2, L1, L2, L3, L4) and 3 experimental halls
(EH1, EH2, EH3).\label{layout}}
\end{figure}
\section{Daya Bay experiment}

The Daya Bay experiment is designed to explore the unknown value of
$\theta_{13}$ by measuring the survival probability of electron
antineutrinos from the nuclear reactors in Daya Bay, China. The Daya
Bay Nuclear Power Plant complex, one of the 5 most prolific sources
of reactor neutrinos in the world, consists of 6 reactors(see
Fig.\ref{layout}) producing 17.4 GW of total thermal power. The goal
of the experiment is to measure $\theta_{13}$ with a sensitivity of
$\sin^{2}2\theta_{13} < 0.01$ at a 90\% C.L. In order to achieve
such high sensitivity, the experiment is optimized in several
aspects. Multiple sites (one far experimental hall (EH3) and two
near experimental halls (EH1,EH2)) are used to effectively cancel
the flux uncertainty by relative measurements. The experiment
employs 8 identically designed detectors to decrease detector
related errors. As shown in Fig.2, only 6 detectors were deployed
before August. 2012. All 8 detectors were installed by October,
2012. The detectors are installed underground to reduce the
cosmic-ray muon flux. Each site has redundant muon detectors (water
Cerenkov detectors and resistive plate chambers (RPCs)) for muon
identification.

\subsection{Detector}

\subsubsection{Antineutrino detector}

\begin{figure}
\centerline{\includegraphics[width=8.0cm]{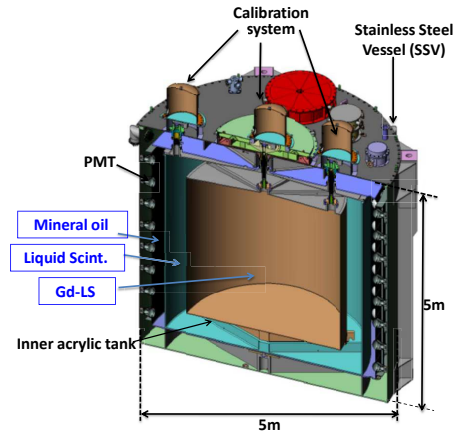}} \vspace*{8pt}
\caption{Antineutrino detector of Daya Bay experiment.\label{AD}}
\end{figure}

The antineutrino detectors (ADs) are filled with Gd doped liquid
scintillator for antineutrino event detection. The experiment uses 8
functionally identical detectors (2 at EH1, 2 at EH2 and 4 at EH3),
which are cylindrical stainless steel vessels(SSV) with a 5 m
diameter and 5 meter height (Fig.\ref{AD}). Each detector is
instrumented with 192 8-inch photomultiplier tubes (PMTs). The
detector has a three-zone structure, including a Gd-doped liquid
scintillator (GdLS) zone, liquid scintillator (LS) zone and mineral
oil (MO) zone. The inner region is the primary target volume filled
with 0.1\% Gd-LS. The middle layer is LS and act as a gamma catcher,
and the outer layer is filled with MO which reduces the impact of
radioactivity from PMT glass and the SSV. Two acrylic tanks are used
to separate each layer. There are two reflectors at the top and
bottom of an AD, which are laminated with ESR (Vikuiti$^{TM}$
Enhanced Specular Reflector Film) film sealed between two 1-cm thick
acrylic panels.The reflectors improve light collection and
uniformity of detector response.

Three automated calibration units (ACUs) are located on the top of
each AD (Fig.\ref{AD}). The ADs are calibrated periodically by
deploying LEDs and radioactive sources inside the detectors. LEDs
are used for timing and PMT gain calibration. Energy calibration is
performed with radioactive sources. Based on the first three months
of data, the first pair of ADs in EH1 are shown to be functionally
identical \cite{side-by-side}.

\subsubsection{Muon veto system}
\begin{figure}
\centerline{\includegraphics[width=7.0cm]{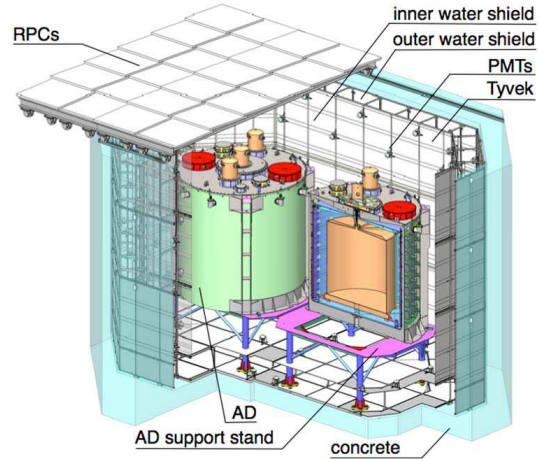}} \vspace*{8pt}
\caption{Veto system of Daya Bay experiment. It includes the inner
water shield, outer water shield and top RPC detector. \label{veto}}
\end{figure}

The muon veto system of  the Daya Bay experiment is shown in
Fig.\ref{veto} . The antineutrino detectors are immersed in an
octagonal pool of ultrapure water. The pool is divided into outer
and inner volumes by Tyvek sheets. Both volumes are instrumented
with PMTs as active muon detectors via muon Cherenkov light.  The
Tyvek sheet with very high reflectivity($>$95\%) can increase light
collection efficiency. The outer layer of the water pool is 1 m
thick and the thickness of the inner layer is $>$1.5 m. At least 2.5
m of water surrounds each AD to shield against ambient
radioactivity. 288 8-inch PMTs are installed in each near hall pool
and 384 in the Far Hall. There is a water circulation and
purification system in each hall to maintain water quality. The tops
of the water Cherenkov detectors are covered by 4 layers of RPCs.
RPCs are gaseous detectors with resistive electrodes operating in
streamer mode at Daya Bay\cite{RPC}. RPC signals are read out by
external strips, with a position resolution of about 8 cm. There are
54 modules in each near hall and 81 modules in the Far Hall. The
designed efficiency is $>$99.5\% with uncertainty less than 0.25\%
by combining the water Cherenkov and RPC detectors. From muon data
analysis, water Cherenkov detector efficiency is $>$99.7\% for long
track muons\cite{side-by-side}, which is better than the design
requirement.

\subsection{Data analysis}
\subsubsection{Event selection}

\begin{figure}
\centerline{\includegraphics[width=8.0cm]{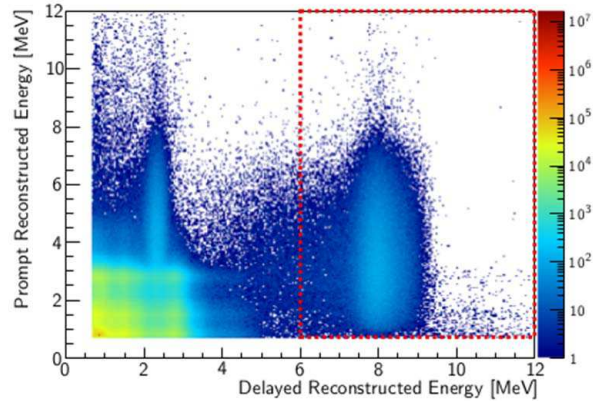}}
\vspace*{8pt} \caption{IBD candidate events are shown within the
dashed lines\label{IBD_evt}}
\end{figure}

  After energy calibration and event reconstruction, we select the IBD events under the following criteria:
\begin{itemlist}
 \item PMT light emission events will be rejected by the PMT flasher cut\cite{side-by-side}.
 \item The energy of prompt and delayed candidates satisfy 0.7 MeV $<$ Ep $<$ 12 MeV and 6.0 MeV $<$ Ed $<$ 12
 MeV, where Ep is the prompt signal energy and Ed is the delayed signal energy.
\item  The delayed signal (neutron capture) time satisfies: 1$\mu$s $<$t $<$
200$\mu$s, where t is the time between the prompt and delayed
 signal.
\item If a muon goes through the water pool and fires $>$12 PMTs, it will
 be treated as a 'water pool muon'. The rejection time window of water pool muons is [-2$\mu$ s, 600$\mu$s].
\item  If the visible energy in the AD is ($>$20 MeV), it will be treated as an 'AD muon'.
 The rejection time window of AD muon is [-2$\mu$s, 1400$\mu$s].
\item  If the visible energy in the AD is $>$2.5 GeV, the rejection time window is [-2 $\mu$ s,0.4s]
\item  There must be no additional prompt-like signals 400$\mu$s before the delayed neutron signal and no additional delayed-like signal 400$\mu$s
after the delayed neutron signal.
\end{itemlist}
As Fig.~\ref{IBD_evt} shows, the events within the dash lines are
the IBD candidate events after selection.

\subsubsection{Background}
   Background control is crucial for a precise measurement in reactor antineutrino experiments. Li9/He8 is
 the main background from cosmic-ray muons. It is a correlated background
 that can mimic prompt signals with a beta-decay signal and delayed signals with neutron
 capture. This background is directly measured by fitting the distribution
  of IBD candidates versus time since the last muon. The background to signal ratio (B/S) of Li9/He8
  is $\sim$0.3$\pm$0.1\%.
  Another background from muons is the fast neutron
 background. Energetic neutrons from muons can give a prompt signal from proton recoil and
 a delayed signal from the neutron's capture on Gd. By relaxing the prompt energy range from 12 MeV to 50 MeV in the IBD events
 selection, the fast neutron
 spectrum in the range of $<$12MeV can be  extrapolated from the higher energy distribution.
 The fast neutron spectrum and rate is cross-checked with fast-neutron samples tagged by the muon veto
system. The B/S ratio of fast neutrons is 0.1\% (0.06\%) for the
near(far) sites.

Two uncorrelated signals can accidentally mimic an IBD signal. This
accident background can be precisely measured from data. The B/S of
accidentals is 1.5\% (4\%) for the near(far) sites. The error in the
accidental rate is 1\% (4\%).

The $^{238}$U, $^{232}$Th, $^{227}$Ac decay chains and $^{210}$Po
produce a $^{13}$C($\alpha$ n)$^{16}$O backgrounds. This background
is estimated by Monte Carlo(MC).The B/S of this background is 0.01\%
(0.05\%) for the near(far) sites. Am-C calibration sources inside
the ACUs on top of each AD constantly emit neutrons, which would
occasionally mimic IBD signals by scattering inelastically with
nuclei in the shielding material (emitting gamma rays) before being
captured on a metal nucleus, such as Fe, Cr, Mn or Ni (releasing
more gamma rays). MC is also used to estimate the rate of this
background. The background rate and shape are constrained by taking
data with a temporary intense source on top of the AD. The B/S ratio
of Am-C sources is 0.04\% (0.35\%) for the near(far) sites.

\subsubsection{Recent results}

\begin{figure}[pb]
\centerline{\includegraphics[width=9.0cm]{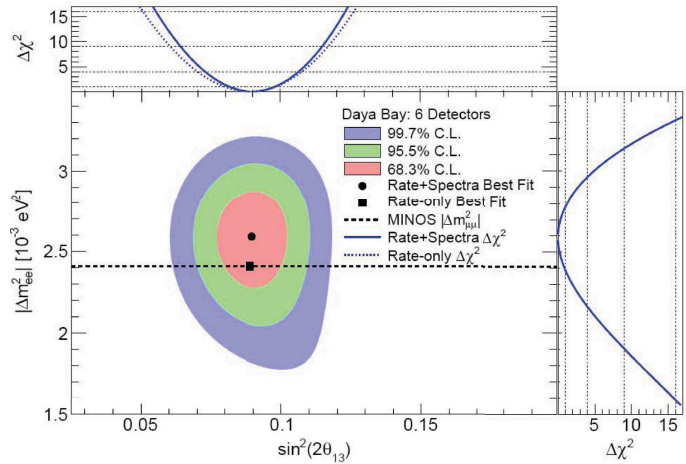}}
\vspace*{8pt} \caption{Allowed regions for the neutrino oscillation
parameters $\sin^{2}(2\theta_{13})$ and $\mid$$\Delta
m^{2}_{ee}$$\mid$ at different confidence levels(solid regions). The
best estimates of the oscillation parameters are shown by the black
dot. The adjoining panels show the dependence of $\Delta \chi^{2}$
on $\mid$$\Delta m^{2}_{ee}$$\mid$(right) and
$\sin^{2}(2\theta_{13})$ (top). The black square and dashed curve
represent the rate-only result. The dashed horizontal line
represents the MINOS $\mid$$\Delta m^{2}_{\mu\mu}$$\mid$
measurement$^{17}$. \label{shape_fit}}

\end{figure}

The Daya Bay experiment first presented the discovery of a non-zero
$\theta_{13}$(5.2$\sigma$) in 2012\cite{DayaBay}. The most recent
analysis\cite{DYBPRL_new} is based on the full six-AD data sample
from Dec. 24, 2011 to July 28, 2012, which includes more than
300,000 IBD interactions. The rate-only analysis yields
$\sin^{2}$$(2\theta_{13})$ = 0.090 $\pm$ 0.010 with $\chiup^{2}$/NDF
= 0.6/4. Spectral information is also used after applying an energy
nonlinearity correction to the positron spectrum. The oscillation
parameters can be extracted from a fit taking into account both rate
and spectral information. In Eq.~\ref{diseq}, the short baseline
($\sim$ km) reactor $\bar \nu_{e}$ oscillation is mainly determined
by the $\Delta_{3i}$ terms. We use the following
definition\cite{mee} of the effective mass-squared difference
$\sin^{2}\Delta ee$ $\equiv$$\cos^2\theta_{12}\sin^{2}\Delta_{31} +
\sin^{2}\theta_{12} \sin^{2}\Delta_{32}$.
 The best-fit values are
$\sin^{2}(2\theta_{13})$ = $0.090^{+0.008}_{-0.009}$ and
$\mid$$\Delta m^{2}_{ee}$$\mid$ = $(2.59^{+0.19}_{-0.20})$$\times
$$10^{-3} \rm eV^{2}$ with $\chiup^{2}$/NDF = 163/153 (68.3\% C.L.). The 68.3\%, 95.5\%, and 99.7\% C.L. allowed
regions in the $\mid$$\Delta m^{2}_{ee}$$\mid$ vs.
$\sin^{2}$$(2\theta_{13})$ plane are shown in  Fig.\ref{shape_fit}.

 The relative deficit and spectral distortion observed
between far and near ADs at Daya Bay give the first independent
measurement of $\mid$$\Delta m^{2}_{ee}$$\mid$ =
$(2.59^{+0.19}_{-0.20})$$\times
$$10^{-3} \rm eV^{2}$ and the most precise measurement of
$\sin^{2}(2\theta_{13})$ = $0.090^{+0.008}_{-0.009}$ to date.

\subsection{Future plan}

\begin{figure}
\centerline{\includegraphics[width=8.0cm]{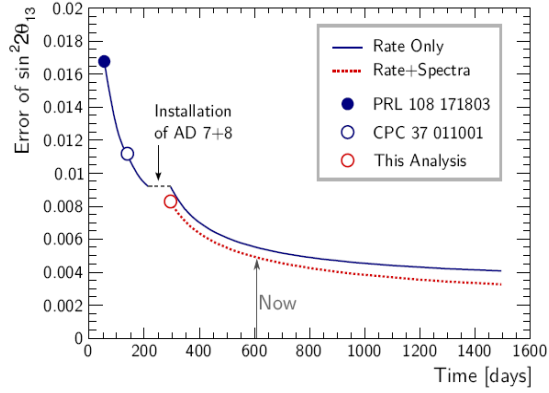}}
\vspace*{8pt} \caption{Error of $\sin^2(2\theta_{13})$ vs time. The
red circle is this analysis result. The current data-taking time is
above 600 days . \label{DYB_sens}}
\end{figure}

\begin{figure}
\centerline{\includegraphics[width=8.0cm]{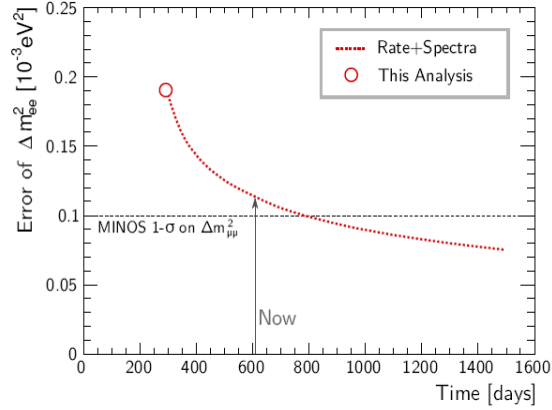}}
\vspace*{8pt} \caption{Error of $\mid$$\Delta m^{2}_{ee}$$\mid$ vs
time. The red circle is this analysis result. The current
data-taking time is above 600 days. The horizontal dashed line is
the error of $\mid$$\Delta m^{2}_{\mu\mu}$$\mid$ from MINOS.
\label{delta_MeeSens}}
\end{figure}

The sensitivity of $\sin^2(2\theta_{13})$ versus time for Daya Bay
is shown in Fig.~\ref{DYB_sens}. The projected uncertainty of
$\sin^2(2\theta_{13})$ is less than 4\% after 3 years of data
taking. Daya Bay is also expecting to measure $\mid$$\Delta
m^{2}_{ee}$$\mid$ in complementary precision to accelerator neutrino
experiments. Fig.\ref{delta_MeeSens} shows the $\mid$$\Delta
m^{2}_{ee}$$\mid$ error versus data-taking time. The error of
$\mid$$\Delta m^{2}_{ee}$$\mid$ will be $<$ $0.10$$\times$$10^{-3}
\rm eV^{2}$ after 3 years of data-taking.

\subsection{RENO and Double Chooz status}

  RENO updated their result in March, 2013\cite{RENO_new}. They gave
 $\sin^{2}(2\theta_{13})$ =0.100$\pm$0.010(stat)$\pm$0.015(syst)
and observed a clear rate deficit (7.1\% reduction) between far and
near sites.  RENO aims to achieve 7\% uncertainty in
$\sin^2(2\theta_{13})$ and suppress their systemic error to 0.5\%.
Double Chooz\cite{Double Chooz_new} provided their new results from
a combined fit of Gd capture and Hydrogen capture with
$\sin^{2}(2\theta_{13})$ =0.109$\pm$0.035. For Double Chooz, the
near detector is under construction and will start running in 2014.

\section{Future prospects}
\begin{figure}
\centerline{\includegraphics[width=8.0cm]{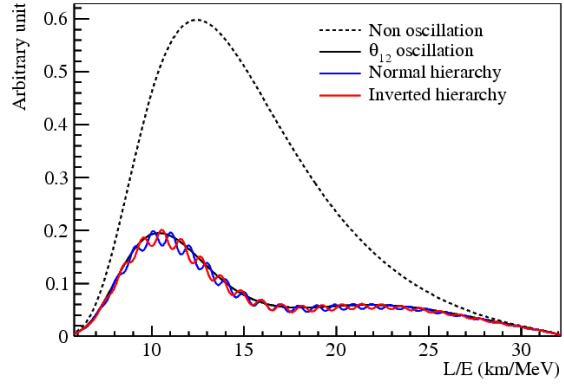}}
\vspace*{8pt} \caption{Neutrino oscillation difference between
normal hierarchy and inverted hierarchy.  \label{MassHer}}
\end{figure}

\begin{figure}
\centerline{\includegraphics[width=8.0cm]{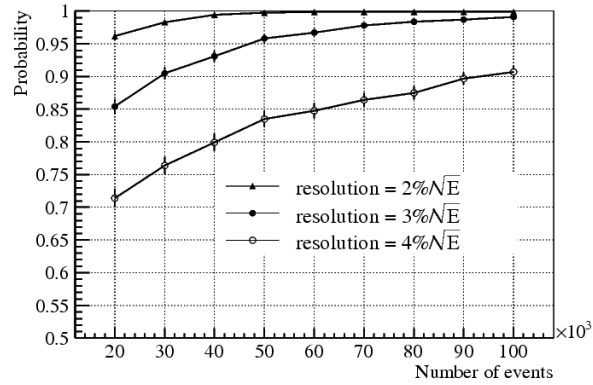}}
\vspace*{8pt} \caption{The probability to determine the sign of the
hierarchy vs. the number of detected events. Different energy
resolutions are used to estimate the effect on the
probability.\label{Mass_sens}}
\end{figure}

The next generation of neutrino oscillation experiments will mainly
focus on the mass hierarchy and CP phase. The mass hierarchy can be
determined by a precisely energy spectrum measurement in reactor
neutrino oscillations, which is independent of the CP-violating
phase. Fig.\ref{MassHer} shows the L/E distribution of reactor
antineutrino oscillations. It is found that frequencies of
oscillation between the normal hierarchy and inverted hierarchy have
some difference. The two oscillation frequency components are driven
by $\Delta m^{2}_{31}$ and $\Delta m^{2}_{32}$ , respectively
\cite{Mass1}$^{-}$\cite{Zhanl1}. The difference is very sensitive to
detector energy resolution. According to the study in Ref.
~\refcite{Zhanl2}, for a single detector at a baseline of 58 km and
with a 35 GW reactor, the probability to determine the sign of the
hierarchy has a significant difference between 3\% and 4\% energy
resolution as shown in Fig.\ref{Mass_sens}. Another big effect on
mass hierarchy sensitivity is potentially from multiple reactor
baselines, which are required to be within 500 m\cite{yufeng}. A
large detector with good energy resolution (3\%) and equal baselines
from powerful reactor cores is required for a mass hierarchy
determination.

\subsection{The Jiangmen Underground Neutrino Observatory}
\begin{figure}
\centerline{\includegraphics[width=9.0cm]{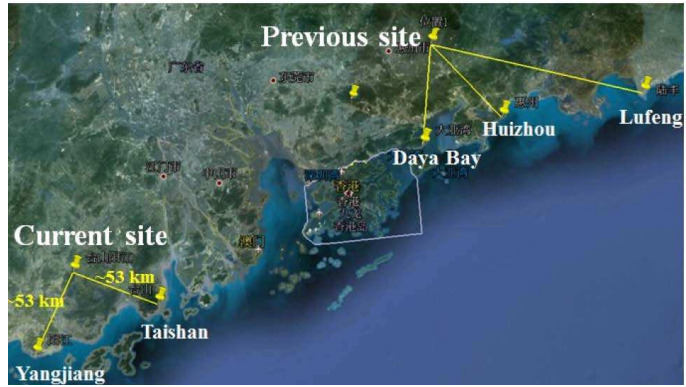}}
\vspace*{8pt} \caption{Experiment site of JUNO. The Taishan and
Yangjiang reactor complexes are used for JUNO. The previous site is
not considered in that the third reactor complex (Lufeng) is being
planed.\label{JUNO_location}}
\end{figure}

 The Jiangmen Underground Neutrino Observatory (JUNO) is located at Kaiping, Jiangmen City, in Guangdong Province, China, as shown in
 Fig.\ref{JUNO_location}, 53 km from the Yangjiang and Taishan nuclear power plants.
 The total thermal power of the reactors is 36 GW.
The detector will be constructed deep underground to reduce the
cosmic-ray muon flux, with an overburden of about 700 m of rock. The
detector will be filled with 20 ktons of liquid scintillator with
3\% energy resolution (at 1MeV). The major goal of JUNO is to
determine the mass hierarchy by precisely measuring the energy
spectrum of reactor antineutrinos. However, it is a multi-purpose
experiment that can also measure neutrino oscillation parameters,
study atmospheric, solar and geo-neutrinos, and perform other exotic
searches.

\begin{figure}
\centerline{\includegraphics[width=9.0cm]{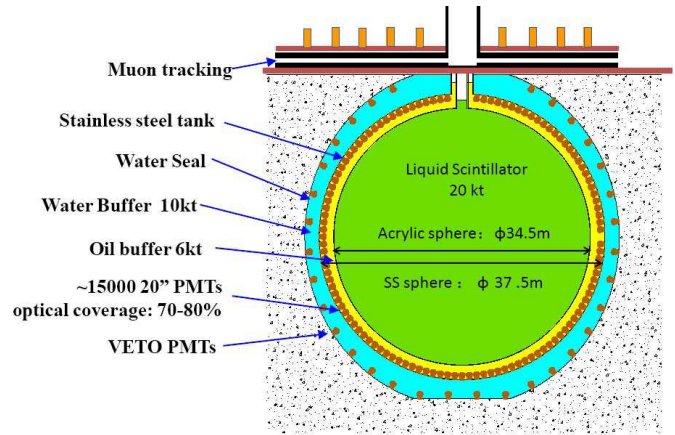}}
\vspace*{8pt} \caption{Detector concept of JUNO, including the
central liquid scintillator detector and outer muon veto systems.
\label{Det_JUNO}}
\end{figure}

 Fig.\ref{Det_JUNO} is the detector concept of JUNO. There are around 15,000 20-inch
PMTs installed in the central detector for the 3\% energy
resolution. 6 ktons of mineral oil is used to shield the
scintillator from PMT glass radioactivity. The muon veto system
includes an outer water Cherenkov detector and a top tracking
system. As for the physics prospects of JUNO\cite{yufeng}, if we
take into account the spread of reactor cores, uncertainties from
energy non-linearity,
 etc, the mass hierarchy sensitivity with 6 years of data-taking can reach $\Delta$$\chi^{2} $$>$ 9 with a relative
 measurement and $\Delta$$\chi^{2}$ $>$16 with an absolute $\Delta m_{\mu\mu}^{2}$ measurement from accelerator neutrino experiments.
 Civil construction has begun and will complete in 2019. Liquid scintillator filling and data taking will begin around 2020.

\subsection{RENO 50}
 RENO-50 was proposed in South
Korea\cite{RENO-50}. The detector will be constructed underground
and consist of 18 ktons of ultra-low-radioactivity liquid
scintillator and 15,000 20-inch PMTs, about 50 km away from the
Hanbit nuclear power plant. The scientific goals of the experiment
include a high precision measurement of $\theta_{12}$ and
$\mid$$\Delta m^{2}_{21}$$\mid$, determination of the mass
hierarchy, the observation of neutrinos from reactors, the Sun, the
Earth, supernovae, and any possible stellar objects. The total
budget is 100 million dollars for 6 years of construction. Facility
and detector construction will be started from 2013 and finished in
2018. The experiment data taking will start in 2019.

\section{Summary}
 Reactor-based antineutrino experiments have obtained many excellent
 achievements in recent years. Daya Bay measured the non-zero $\theta_{13}$ with great precision, together
  with other experiments. The most recent results from Daya Bay provide the first independent measurement of
$\mid$$\Delta m^{2}_{ee}$$\mid$ and the most precise measurement of
$\sin^{2}(2\theta_{13})$ to date. The precision on
$\sin^{2}(2\theta_{13})$ will be improved to ~4\% in the future. The
large value of $\theta_{13}$ boosts the next generation of neutrino
oscillation experiments to determine the neutrino mass hierarchy and
measure the CP violation phase. Reactor-based antineutrino
experiments will continue to play an important role in the mass
hierarchy determination and precise measurements of oscillation
parameters.

\section*{Acknowledgments}
 The article is supported by the National Natural Science Foundation of China
(Y3118G005C). I would like to acknowledge my Daya Bay collaborators
for useful suggestion and comments, especially Dr. Yufeng Li, Logan
Lebanowski and Viktor Pec who helped me a lot to improve this
article.


\end{document}